\documentclass[]{achemso}

\usepackage{graphicx}
\usepackage{natbib}
\usepackage{epstopdf}
\usepackage{dcolumn}
\usepackage{bm}
\usepackage{color}
\usepackage[symbol*]{footmisc}

\title{Anisotropic Pauli Spin Blockade of Holes in a GaAs Double Quantum Dot}

	\author{Daisy Q. Wang}
	\email{qingwen.wang@unsw.edu.au}
	\author{Oleh Klochan}
	\author{Jo-Tzu Hung}
	\author{Dimitrie Culcer}
	\affiliation{School of Physics, University of New South Wales, Sydney NSW 2052, Australia}
    \author{Ian Farrer}  
    \affiliation{Cavendish Laboratory, J. J. Thomson Avenue, Cambridge, CB3 0HE, United Kingdom} 
    \altaffiliation{Present Address: Department of Electronic and Electrical Engineering, The University of Sheffield, Mappin Street, Sheffield, S1 3JD, United Kingdom.}
	\author{David A. Ritchie}
	\affiliation{Cavendish Laboratory, J. J. Thomson Avenue, Cambridge, CB3 0HE, United Kingdom}
	\author{Alexander R. Hamilton}
	\email{alex.hamilton@unsw.edu.au}
	\affiliation{School of Physics, University of New South Wales, Sydney NSW 2052, Australia}

\begin{document}
	
	This document is the unedited Author's version of a Submitted Work that was subsequently accepted for publication in Nano Letters, copyright~\textsf{\copyright} American Chemical Society after peer review. To access the final edited and published work see
	http://pubs.acs.org/doi/full/10.1021/acs.nanolett.6b03752.
	
\begin{abstract}
	Electrically defined semiconductor quantum dots are attractive systems for spin manipulation and quantum information processing. Heavy-holes in both Si and GaAs are promising candidates for all-electrical spin manipulation, owing to the weak hyperfine interaction and strong spin-orbit interaction. However, it has only recently become possible to make stable quantum dots in these systems, mainly due to difficulties in device fabrication and stability. 
	Here we present electrical transport measurements on holes in a gate-defined double quantum dot in a $\mathrm{ GaAs/Al_xGa_{1-x}As}$ heterostructure. We observe clear Pauli spin blockade and demonstrate that the lifting of this spin blockade by an external magnetic field is highly anisotropic. Numerical calculations of heavy-hole transport through a double quantum dot in the presence of strong spin-orbit coupling shows quantitative agreement with experimental results and suggests that the observed anisotropy can be explained by both the anisotropic effective hole g-factor and the surface Dresselhaus spin-orbit interaction.
\end{abstract}


Recently, all-electrical control of single electron spins has been demonstrated in electron systems with strong spin-orbit coupling using electric dipole spin resonance (EDSR) techniques~\cite{NadjPerge10, Berg13}. 
Utilizing the coupling between spin and orbital states, an oscillating electric field can effectively rotate the electron spin coherently~\cite{Flindt06}. 
However, most electron systems with spin-orbit coupling also have a significant hyperfine interaction with the nuclei in the host crystal~\cite{Pfund07,Schroer11,NadjPerge102}. This electron-nuclear spin interaction causes unavoidable spin dephasing, and is the dominant factor limiting the spin lifetimes~\cite{Jouravlev06}. Valence-band holes also have strong spin-orbit coupling, but have much weaker hyperfine interaction with nuclear spins due to the p-orbital symmetry of their Bloch wavefunction~\cite{Bulaev05,Fischer08,Fischer10,Chekhovich13}. 
Therefore, hole spins have drawn significant attention recently as a possible solution to improve the spin lifetimes for all-electrical spin manipulation. 
Nonetheless, understanding of spin properties of holes in quantum dots is still limited, and to date there have been few studies of spin-dependent electrical transport in hole quantum dots~\cite{Pribiag13,Higginbotham14,Li15}. 
Because of the larger effective mass of holes compared to electrons, hole quantum dots need to have much smaller dimensions to observe transport through orbital states in the few-hole limit. One approach is to use nanowire-based few-hole quantum dots~\cite{Higginbotham14,Pribiag13}, but these have light-hole ground states. In contrast, quantum dots formed by surface gates on a 2D heterostructure should have heavy-hole characteristics. In addition, surface-gate-defined lateral quantum dots are more amenable to scale-up to multiple dots for complex qubit operations~\cite{Takakura14,Ward16}.

Pauli spin blockade is a simple and effective tool for detecting spin dependent transport in an all-electrical measurement, and forms the basis of many advanced spin manipulation and quantum information processing experiments. Even though spin blockade has been widely observed in spin-1/2 (electron and light-hole) systems~\cite{Hanson07}, to the best of our knowledge, it has not been demonstrated in GaAs-based spin-3/2 heavy-hole systems. 
In this paper, we report the observation of Pauli spin blockade of holes, for the first time, in electrical transport measurements of a gate-defined double quantum dot on a GaAs/Al$_x$Ga$_{1-x}$As heterostructure. By applying an external magnetic field in different directions, we study the anisotropic lifting of this spin blockade due to spin-orbit coupling.

\begin{figure*}
	\centering
	\includegraphics[width=0.9\linewidth]{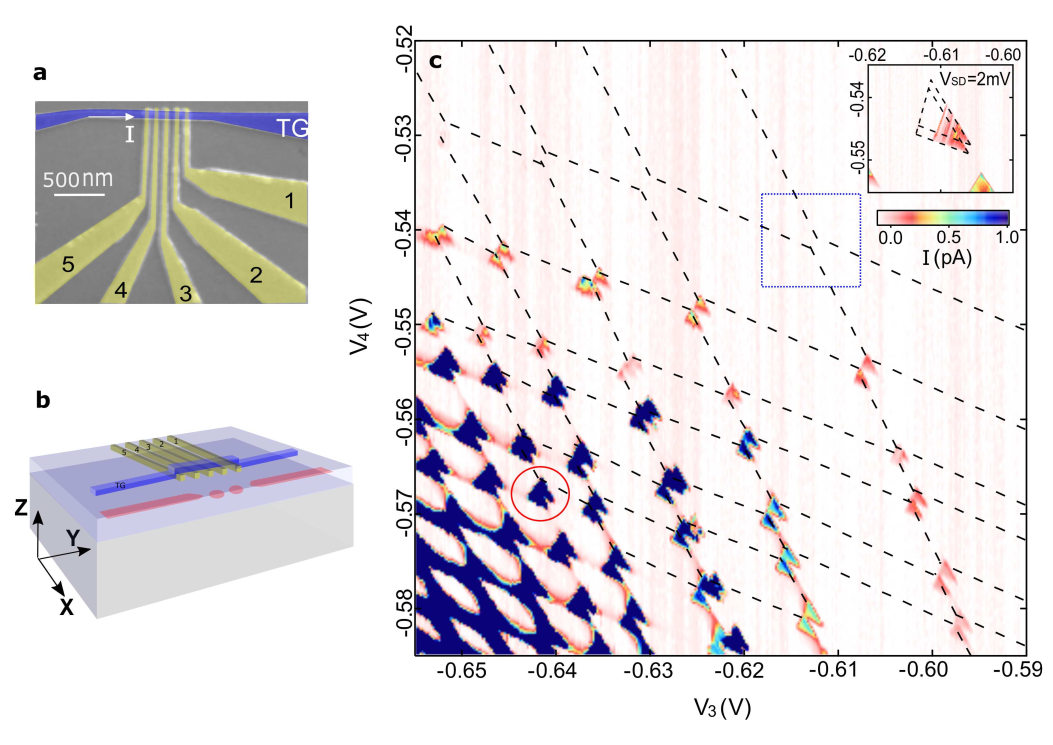}
	\caption{\small 
			(a) A false-coloured Scanning Electron Microscope (SEM) image of the device. The first layer of gates consists of five electrodes (yellow) with a width of 30 nm and a inter-gate spacing of 50 nm. A top channel gate (blue) on the top layer has a width of 50 nm. 10 nm of $\mathrm{HfO_x}$ is used as the insulator between the two layers of Ti/Au gates; (b) A 3D schematic of the device. The undoped $\mathrm{GaAs/Al_xGa_{1-x}As}$ heterostructure has a 10 nm GaAs cap and 50 nm $\mathrm{Al_xGa_{1-x}As}$ layer. To tune the device into a double quantum dot, the top-gate is negatively biased to V$_{\mathrm{TG}}=-1.05$ V to induce holes at the heterointerface. Gates 2 and 5 are used as the left and right barriers of the dot, while gate 1 is not used (kept at V$_1=-0.73$ V as part of the lead). Gates 3 and 4 are used as plunger gates for left and right dots respectively, and control the inter-dot coupling at the same time. The two quantum dots (red) are confined at the heterointerface 60nm from the top of the wafer. (c) Charge stability diagram of the double quantum dot: current through the dot measured as a function of the voltages on the left plunger (gate 3) and the right plunger (gate 4) with V$_{\mathrm{SD}}=0.5$ mV. Dashed lines are guides to the eye outlining the typical honeycomb pattern for a double quantum dot. Inset: Charge stability diagram of the last visible pair of bias triangles (highlighted by the rectangle) with V$_{\mathrm{SD}}=2$ mV.}
	\label{fig:device3}
\end{figure*}

{\bf The operation of a few-hole double quantum dot.}
To reach the few-hole limit we use a quantum dot device with the double-layer-gate design~\cite{Wang16} shown in Figs~\ref{fig:device3}(a) and (b). 
The operation of the double dot is demonstrated by the charge stability diagram shown in Fig~\ref{fig:device3}(c). The size of the honeycombs increases rapidly as gate biases V$_3$ and V$_4$ are made more positive, which suggests the dots are in the few-hole regime.
The measured addition energy of E$_{\mathrm{add}}=3-4$ meV for the second hole in both dots is also comparable to measurements with the same device in a single-dot configuration. For the last row of bias triangles the current is very small due to imbalanced tunnel barriers. We plot in the inset of Fig~\ref{fig:device3}(c) the last observable pair of bias triangles with V$_{\mathrm{SD}}=2$ mV. The strong suppression of I$_{\mathrm{SD}}$ in the base of these triangles explains why they were not visible with V$_{\mathrm{SD}}=0.5$ mV.

\begin{figure*}
	\centering
	\includegraphics[width=0.9\linewidth]{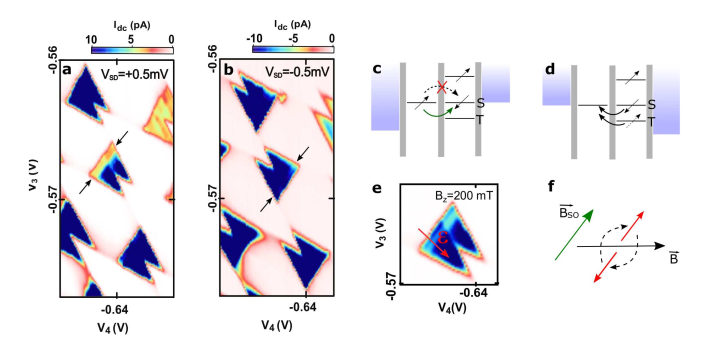}
	\caption{\small{\bf Pauli spin blockade of holes.} Stability map focusing on the pair of bias triangles identified by the red circle in Fig~\ref{fig:device3}(c), showing signatures of Pauli spin blockade indicated by the black arrows: (a) with V$_{\mathrm{SD}}=+0.5$ mV, and (b) with V$_{\mathrm{SD}}=-0.5$ mV. (c) and (d) are schematics showing the equivalent charge transport through the dot with positive and negative biases respectively. (e) Stability map of the spin blocked pair of bias triangles at $B_z=200$ mT with a source-drain bias of V$_{\mathrm{SD}}=0.5$ mV. The magnetic field lifts the spin blockade in the base of the bias triangle. The red arrow indicates the direction of the detuning axis $\varepsilon$. (f) Schematic showing the rotation of the hole spin (red arrows) in an external magnetic field $\vec{B}$ due to spin-orbit interaction.}
	\label{fig:spinblockade3}
\end{figure*}

{\bf Pauli spin blockade.}
Signatures of Pauli spin blockade are observed in several pairs of bias triangles with different hole occupations, and here we focus on the bias triangles highlighted by the red circle in Fig~\ref{fig:device3}(c).
Figs~\ref{fig:spinblockade3}(a) and (b) show a zoom-in of the region around these bias triangles for positive and negative V$_\mathrm{SD}$. Comparing the two figures, the top and bottom pairs of bias triangles look very similar, whereas the current through the base of the middle pair of bias triangles (highlighted by the black arrows) is strongly suppressed in the positive-bias direction but flows freely in the negative-bias direction. This is a characteristic signature of Pauli spin blockade, as illustrated in the schematics in Fig~\ref{fig:spinblockade3}(c) and (d). Current is suppressed in the positive-bias direction as transport through the only energetically allowed S(0, 2) singlet state is blocked when a T(1, 1) triplet state is occupied. The bias at which Pauli spin blockade is lifted gives the singlet-triplet splitting $\Delta_{\mathrm{ST}}\sim$150 $\mu$eV. 

\begin{figure*}
	\centering
	\includegraphics[width=0.9\linewidth]{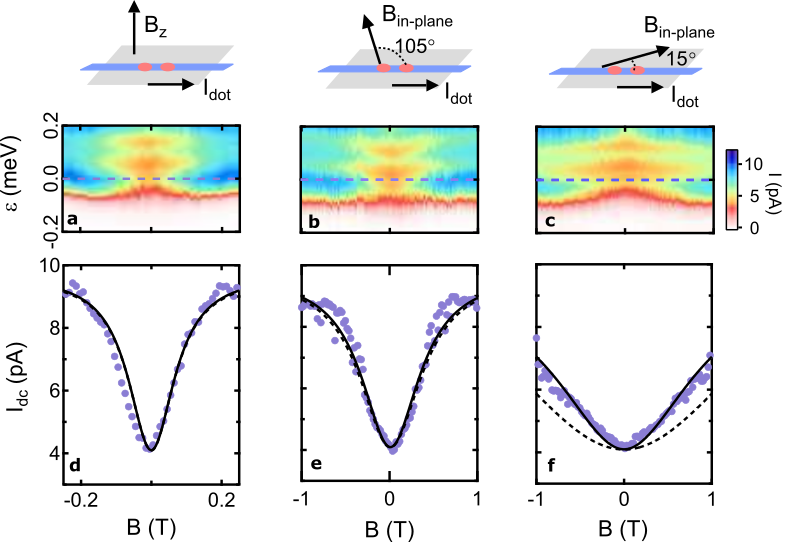}
	\caption{\small{\bf Lifting of Pauli spin blockade in a magnetic field.} Current through the double dot plotted as a function of the detuning energy and magnetic field along different directions. Dashed lines indicate the linecuts at zero detuning. The data is taken by applying V$_{SD}=0.5$ mV and sweeping the right plunger gate voltage V$_4$ along a linecut at V$_3=-0.5676$ V while stepping the magnetic field (a) out-of-plane, (b) in-plane $105^{\circ}$ from the double-dot axis and (c) in-plane $15^{\circ}$ from the double-dot axis (due to a small misalignment between the magnets and the sample). (d)-(f) current through the dot at zero detuning (the linecuts through the data in (a)-(c) shown by the dashed lines) as a function of magnetic field for the three different field directions. Solid (and dashed) lines show numerical calculations of the dot current with (and without) B-dependent spin relaxation.}
	\label{fig:Bxyz3}
\end{figure*}

In addition to the dependence of I$_{\mathrm{SD}}$ on bias direction, another signature of Pauli spin blockade is the effect of a magnetic field. As shown in Fig~\ref{fig:spinblockade3}(e), applying a small perpendicular magnetic field of B$_z=200$ mT almost fully recovers the current through the spin blocked region. 
In GaAs hole systems, the strong spin-orbit interaction (SOI) leads to hybridization of the $T(1, 1)$ triplet and the $S(2, 0)$ singlet states, which allows the previously forbidden $T(1, 1)\to S(2, 0)$ transition and lifts the spin blockade at finite B. In the simple physical picture shown in Fig~\ref{fig:spinblockade3}(f), spins are oriented along the intrinsic effective spin-orbit field direction $\vec{B}_{\mathrm{SO}}$, so that an external magnetic field $\vec{B}$ applied perpendicular to $\vec{B}_{\mathrm{SO}}$ causes the spin to precess around $\vec{B}$. This rotates the spin and enables spin-flip tunnelling to lift the spin blockade. 
The detailed response of the leakage current induced by SOI is shown in Fig~\ref{fig:Bxyz3}(a) as a function of detuning and B$_z$, and a linecut at zero-detuning is plotted in Fig~\ref{fig:Bxyz3}(d). 
The current in the spin-blocked region increases monotonically as the field increases, until it reaches the value of the non-blocked case.

A more interesting situation is when an in-plane magnetic field is applied, since the orientation of $\vec{B}_{\mathrm{SO}}$ 
always points in the plane of charge motion and depends strongly on the nature of the dominant SOI. For a heavy-hole system, Rashba SOI creates an effective $\vec{B}_{\mathrm{SO}}$ perpendicular to $\vec{k}$, which can be simply considered as the current direction or the double-dot axis. On the other hand, if Dresselhaus SOI is dominant, $\vec{B}_{\mathrm{SO}}$ depends on the crystalline orientation and is parallel to $\vec{k}$ for heavy-holes in (100) GaAs (see Supplementary information S3). Therefore, to gain more information about the SOI in our device, we plot in Figs~\ref{fig:Bxyz3}(b) and (c) the leakage current when an in-plane magnetic field is applied along two orthogonal directions. Even though a similar zero-field dip is observed for both cases, the widths of the dip are dramatically different. A full recovery of the spin-blocked current occurs at $B\sim0.8$ T when the field is roughly perpendicular to the double-dot axis, whereas when the magnetic field is applied roughly parallel to the double-dot axis, no saturation of dot current is observed up to 1 T.

\begin{figure}
	\centering
	\includegraphics[width=0.8\linewidth]{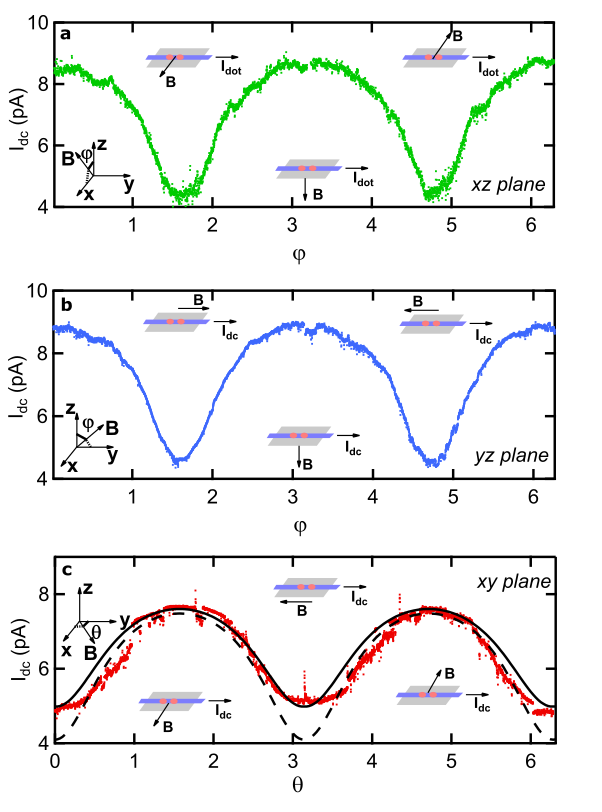}
	\caption{\small{\bf Anisotropic lifting of Pauli spin blockade.} Current through the double dot as a function of $\varphi$ (the angle between $\vec{B}$ and the $z$-axis) and $\theta$ (the angle between $\vec{B}$ and the $y$-axis in the $xy$ plane) while rotating a fixed magnetic field (a) $B=0.2$ T in $xz$ plane, (b) $B=0.2$ T in $yz$ plane and (c) $B=0.5$ T in $xy$ plane. Solid and dashed black lines plot the numerical calculation of the dot current with and without including B-dependent spin relaxation processes.}
	\label{fig:Brot}
\end{figure}

To systematically investigate the anisotropy of the lifting of spin blockade, a fixed magnetic field is rotated in the $xz$, $yz$ and $xy$ planes, while monitoring the current through the double dot in the spin-blocked regime. Figs~\ref{fig:Brot}(a) and Fig~\ref{fig:Brot}(b) show the leakage current at zero-detuning as a fixed magnetic field $B=0.2$ T is rotated in the $xz$ and $yz$ planes, and Fig~\ref{fig:Brot}(c) shows what happens when the magnetic field is rotated in the $xy$ plane (using a slightly larger field of 0.5 T). 

As most theoretical studies of Pauli spin blockade have focussed on spin-1/2 electrons~\cite{Danon09,NadjPerge102}, to understand the anisotropy observed in our system, we follow the approach in Ref.~\citenum{Danon09} and calculate heavy-hole transport through a double quantum dot in the presence of strong SOI. We start with a $4\times4$ Hamiltonian including Zeeman, Dresselhaus and Rashba SOI terms. Assuming heavy-hole light-hole mixing is small enough to be considered as a perturbation, we numerically evaluate the spin-conserving and spin-flipping tunnelling matrix elements between all (1, 1) states and the (0, 2) singlet. The resulting current through the double dot is dependent on both the hole g-factor (through the Zeeman effect) and the orientation of the external magnetic field $\vec{B}$ relative to $\vec{B}_{\mathrm{SO}}$~\cite{Jo} (see Supplementary information S1 and S2). The measured current can be modelled with an interdot tunnel coupling $t_0=200$ $\mu$eV, a ratio $t_{\mathrm{SO}}/t_0=0.34$ between spin-flipping and spin-conserving tunnelling processes, and a spin relaxation rate $\Gamma_{\mathrm{rel}}=3.2$ MHz, which is comparable to previous double-dot measurements~\cite{NadjPerge102,Li15}. The calculated current shows good qualitative agreement with the measurement, as depicted by the black lines in Figs~\ref{fig:Bxyz3}(d) and (e). 
A small discrepancy between the calculation and the experiment appears when the magnetic field is almost parallel to the double-dot axis, shown by the dashed lines in Fig~\ref{fig:Bxyz3}(f) and Fig~\ref{fig:Brot}(c). This small discrepancy can be accounted for by including a B-dependent spin relaxation process ($\propto$B$^2$) due to piezoelectric phonon coupling~\cite{Baruffa10,Bulaev05,Bulaev07} (see Supplementary information S4). Fitting the data (solid line in Fig~\ref{fig:Bxyz3}(f)) yields a B$^2$-dependent spin relaxation rate of $\Gamma_B\sim0.2$ GHz at $B=1$ T. This implies a rather strong hole-phonon coupling, consistent with previous measurement of holes in GaAs~\cite{Gao05}. Note that SO-assisted relaxation due to phonon coupling is only visible when $\vec{B}$ is aligned along the double-dot axis, as it only enhances the leakage current when the tunnelling rate induced by SOI is slow. 

From the theoretical simulations the extreme anisotropy of the leakage current in Fig~\ref{fig:Brot} can be understood as arising from two processes: firstly, when the magnetic field is tilted out-of-plane the Zeeman splitting between triplets increases dramatically due to the highly anisotropic heavy-hole g-factor in GaAs heterostructures~\cite{Winkler,Simion14}, which causes the anisotropic leakage current in Figs~\ref{fig:Brot}(a) and (b). 
Secondly, when the magnetic field is varied in-plane (Fig~\ref{fig:Brot}(c)), the relative orientation of $B_{\parallel}$ with respect to $\vec{B}_{\mathrm{SO}}$ changes, which affects the efficiency of the spin-flip tunnelling process. One extreme case is when $B_{\parallel}$ is aligned with $\vec{B}_{\mathrm{SO}}$. In this case, no spin-flip tunnelling can be induced by SOI, so spin blockade persists and the current remains suppressed.
Surprisingly, the minimum current of our device is observed when $B_{\parallel}$ is applied $along$ the double-dot axis, which indicates that $\vec{B}_{\mathrm{SO}}$ is parallel to the dot current. This result is very different from measurements on electron systems with strong SOI, where the suppression of current is observed when $B_{\parallel}$ is applied $perpendicular$ to the double-dot axis~\cite{NadjPerge12}. Our result is also distinct from previous studies of light-holes in nanowire dots, where no strong dependence of the current on the orientation of $B_{\parallel}$ was observed~\cite{Pribiag13}. The difference in the orientation of $\vec{B}_{\mathrm{SO}}$ between electrons, light-holes and heavy-holes highlights the fundamental differences between spin-1/2 and spin-3/2 systems. One possible explanation for the orientation of $\vec{B}_{\mathrm{SO}}$ observed here is that the Dresselhaus SOI is much stronger than the Rashba SOI. This could be caused by the surface Dresselhaus SOI, which has been shown to be much larger than bulk Dresselhaus at a heterointerface~\cite{Durnev14}. Alternatively, the orientation of $\vec{B}_{\mathrm{SO}}$ could also be varied by transport through higher orbital excited states in elliptical quantum dots, 
although these should be energetically suppressed and in this case there would be no reason why $\vec{B}_{\mathrm{SO}}$ should be aligned with the double-dot axis.

In conclusion, we present measurements of hole spin blockade in a double quantum dot. We observe a large increase in the leakage current when an external magnetic field is applied, which suggests the lifting of spin blockade due to SOI. By varying the magnetic field orientation, we demonstrate the anisotropic behaviour of the lifting of spin blockade. Intriguingly this anisotropy is very different to that observed for both electrons and light-holes. Numerical calculations yield quantitative agreement with experimental results, and suggest that the observed anisotropy can be due to a combination of the anisotropic hole g-factor and the Dresselhaus SOI. 



{\large{\bf Acknowledgements}}

This work was funded by the Australian Research Council under the DP and DECRA schemes, and the EPSRC (UK). Devices were made at the NSW node of the Australian National Fabrication Facility. We thank O. Sushkov, D. Miserev, M. V. Durnev, L. E. Golub and A. Sachrajda for helpful discussions.

{\large{\bf References}}
\bibliographystyle{plainnat}






\end{document}